%% file: main.tex
\documentclass[10pt,conference]{IEEEtran}
\IEEEoverridecommandlockouts
\usepackage{cite}
\usepackage{xcolor}
\usepackage{amsmath}
\usepackage{amssymb}
\usepackage{amsthm}
\usepackage{amsfonts}
\usepackage{algorithm}
\usepackage{algpseudocode}
\usepackage{import}
\usepackage{braket}
\usepackage{qcircuit}
\usepackage{adjustbox}
\usepackage{graphicx}
\usepackage{balance}
\usepackage{multirow}
\usepackage{hyperref}
\usepackage{booktabs}

\newcommand{\fig}{Fig.~}

\def\BibTeX{{\rm B\kern-.05em{\sc i\kern-.025em b}\kern-.08em
    T\kern-.1667em\lower.7ex\hbox{E}\kern-.125emX}}
\begin{document}

\title{Memory-, Circuit-, and Ansatz-Efficient VQLS for CFD on Hybrid Quantum–HPC Systems}

\author{$^\dagger$Chao Lu, $^\dagger$Muralikrishnan Gopalakrishnan Meena, Eduardo Antonio Coello P\'{e}rez,\\ Kalyana Chakravarthi Gottiparthi, and Seongmin Kim\\

\IEEEauthorblockA{National Center for Computational Sciences, Oak Ridge National Laboratory, Oak Ridge, TN, USA\\}
\IEEEauthorblockA{$^\dagger$These authors contributed equally, Correspondence: \href{mailto:luc1@ornl.gov}{luc1@ornl.gov}, \href{mailto:gopalakrishm@ornl.gov@ornl.gov}{gopalakrishm@ornl.gov@ornl.gov}}}

\maketitle

\begin{abstract}
\import{Sections/}{Abstract}\footnote{\textbf{Notice:} This manuscript has been authored by UT-Battelle, LLC, under contract DE-AC05-00OR22725 with the US Department of Energy (DOE). The US government retains and the publisher, by accepting the article for publication, acknowledges that the US government retains a nonexclusive, paid-up, irrevocable, worldwide license to publish or reproduce the published form of this manuscript, or allow others to do so, for US government purposes. DOE will provide public access to these results of federally sponsored research in accordance with the DOE Public Access Plan (\href{http://energy.gov/downloads/doe-public-access-plan}{http://energy.gov/downloads/doe-public-access-plan}).}
\end{abstract}


\import{Sections/}{Introduction}
\import{Sections/}{Related_Works}
\import{Sections/}{Methodology}
\import{Sections/}{Results}
\import{Sections/}{Conclusion}
\import{Sections/}{Acknowledgments}

\balance
\bibliographystyle{IEEEtran}
\bibliography{references}
\end{document}

%% file: Sections/Abstract.tex
Fluid dynamics workloads are dominated by repeated solves of large, structured linear systems, motivating the search for quantum acceleration. The Variational Quantum Linear Solver (VQLS) is a leading near-term candidate, but practical deployment on hybrid quantum--high--performance computing (HPC) systems faces three persistent challenges: (i) the linear-combination-of-unitaries (LCU) encoding of the system matrix explodes in memory and runtime as the problem size grows, (ii) ansatz selection is largely empirical, with no clear link between standard circuit metrics and solver convergence, and (iii) end-to-end VQLS pipelines have rarely been exercised on production HPC hardware at non-trivial qubit counts.

This work addresses these challenges through three contributions. First, we benchmark four matrix-encoding strategies---naive LCU, PennyLane-integrated, Fast Walsh--Hadamard Transform (FWHT)-based parallel Pauli decomposition, and an singular value decomposition (SVD)-based two-term LCU---and show that the FWHT approach reduces peak memory by up to $1298\times$ on an $11\times 11$ Hele--Shaw grid, while the SVD-based coherent VQLS delivers over $10{,}000\times$ per-iteration speedup over standard Pauli-based VQLS at 8 qubits. Second, we evaluate 11 ansatz families with gradient-free and gradient-based optimizers on canonical Hele--Shaw flow, and find that expressibility and entanglement metrics correlate only weakly with VQLS convergence, motivating problem-aware ansatz design. Third, we deploy the full workflow on the OLCF Frontier supercomputer and successfully simulate a 15-qubit tridiagonal Toeplitz system on a single node. Together, these results establish a practical baseline for VQLS in hybrid quantum--HPC computation fluid dynamic (CFD) workflows and identify the remaining bottlenecks for larger problems.

%% file: Sections/Introduction.tex
\section{Introduction} \label{sec:intro}

Solving systems of linear equations of the form $A\mathbf{x}=\mathbf{b}$ is a fundamental building block of scientific computing. It appears as a subroutine in diverse workloads such as regression and kernel methods in machine learning, large-scale optimization, uncertainty quantification, and the simulation of complex physical phenomena ranging from transport processes to fluid dynamics and weather prediction~\cite{meanti2020kernel, travelletti2023uncertainty, demmel1997applied, clancy2013use}. In many of these settings, the linear system does not arise in isolation: it is repeatedly solved inside nonlinear iterations, time-stepping loops, or parameter sweeps. Consequently, even moderate improvements in per-solve cost can translate into substantial end-to-end savings in wall-clock time and energy consumption on modern high-performance computing (HPC) platforms.


From the classical perspective, the cost of solving $A\mathbf{x}=\mathbf{b}$ depends strongly on properties of the matrix $A$, including the sparsity $s$, the dimension $N$, and the condition number $\kappa$. While dense direct methods scale superlinearly in $N$, state-of-the-art solvers exploit sparsity and structure through iterative methods and preconditioning. Nevertheless, the overall complexity can remain high, especially when many right-hand sides, multiple parameter points, or stringent accuracy requirements are involved. At a coarse level, classical algorithms for linear systems exhibit computational costs that can be expressed as $\mathcal{O}(N^m)$ with $m\in[1,2,3]$ depending on the algorithmic regime and matrix structure~\cite{coppersmith1982asymptotic,trefethen2022numerical}. These scaling considerations, together with the growing demand for larger simulations and higher fidelity models, motivate the search for alternative computing approaches such as quantum computing, which has emerged as a promising candidate for accelerating linear algebra workloads through quantum linear systems algorithms (QLSAs) and, more recently, variational quantum linear solvers (VQLS) tailored to NISQ-era hardware (reviewed in Section~\ref{sec:related}).

Motivated by the challenges of deploying VQLS for realistic CFD-style problems, this paper aims to provide a workflow-oriented view of VQLS for CFD-relevant linear systems: not only whether VQLS can solve a given small instance, but how different implementation choices affect accuracy, circuit cost, and scalability trends as the problem size grows. We focus on canonical, structured test problems that are representative of fluid-flow discretizations while allowing controlled scaling studies. We also emphasize the algorithm engineering aspects that are essential for integration into HPC systems, where throughput, parallel execution, and reproducibility are as important as single-instance performance.

Our contributions are:
\begin{itemize}

    \item We develop and benchmark an end-to-end VQLS workflow, from matrix encoding through optimization, using canonical test problems (tridiagonal Toeplitz and Hele--Shaw flow) as validation targets and performed simulation on the Frontier supercomputer.
    \item We implement and compare multiple linear-combination-of-unitaries (LCU) decomposition methods for VQLS initialization and quantify their runtime and memory trade-offs.
    \item We introduce an approximate (tolerance-controlled) parallel Pauli decomposition and characterize its accuracy--cost trade-off on large Toeplitz systems, demonstrating how truncation can drastically reduce LCU term count and VQLS circuit workload while controlling the relative Frobenius error.
    \item We evaluate 11 quantum ansatz families on a canonical Hele--Shaw problem and connect convergence behavior to expressibility and entanglement metrics.

\end{itemize}

The rest of the paper is organized as follows: Section~\ref{sec:related} reviews related work on quantum linear systems algorithms and VQLS, and identifies the open challenges that motivate our study. Section~\ref{sec:Methodology} provides the mathematical background for VQLS and the target CFD problems and the detail of our implementation of matrix decompositions and the HPC-integrated workflow of VQLS and its variants. Section~\ref{sec:results} presents our numerical results, including the ansatz comparison and scaling studies on the Frontier supercomputer. Finally, Section~\ref{sec:Conclusion} concludes with a discussion of future directions for quantum linear solvers and potential applications.

%% file: Sections/Related_Works.tex
\section{Related Works} \label{sec:related}

Quantum computing introduced the possibility of asymptotic speedups for certain linear algebra tasks. In particular, quantum linear systems algorithms (QLSAs) aim to prepare a quantum state proportional to the solution vector, $\ket{x}\propto A^{-1}\ket{b}$, rather than outputting all components of $\mathbf{x}$ explicitly. The Harrow--Hassidim--Lloyd (HHL) algorithm was the first widely recognized QLSA and suggested an exponential improvement in the dependence on $N$~\cite{harrow2009quantum}. However, HHL and related approaches typically incur polynomial overhead in $\kappa$ and $s$, and they rely on quantum phase estimation (QPE), which can increase circuit depth tremendously as tighter approximation error $\epsilon$ and accurate encoding of the input matrix are required. These factors are critical: for matrices encountered in real-world discretized PDEs, the condition number can grow rapidly with mesh refinement, and deep circuits are challenging on current devices.

Numerous variations of the HHL algorithm have been introduced in the past decade to reduce the dependence on $\kappa$ and QPE \cite{ambainis2012variable,childs2017quantum,subacsi2019quantum,an2022quantum,lu2025lugo,morales2024quantum}, with the best approach demonstrating a complexity of $\mathcal{O}(\kappa \log{\kappa/\epsilon})$\cite{lin2020optimal,costa2022optimal,jennings2023efficient}. However, the large circuits generated from these algorithms limit their scalability to practical problems on current Noisy Intermediate-Scale Quantum (NISQ) computers. Practical demonstrations of such algorithms on current hardware remain limited to small or highly structured instances. In the current era of NISQ devices, variational quantum--classical algorithms are appealing for practical workloads\cite{cerezo2021variational}. One such variational algorithm for solving linear systems of equations was introduced by \cite{bravo2023variational}, namely the Variational Quantum Linear Solver (VQLS).

While fault-tolerant quantum computers remain a long-term goal, the present generation of NISQ devices has encouraged the development of hybrid quantum--classical algorithms that can operate with limited qubit counts and constrained circuit depths. Variational approaches can replace deep QPE subroutines with parameterized quantum circuits trained by a classical optimizer~\cite{cerezo2021variational}. VQLS follows this philosophy by formulating the linear solve as an optimization problem and by leveraging measurements to evaluate a cost function whose minimizer corresponds to the desired solution state.

VQLS has been explored as a route to scalable quantum linear solvers on existing hardware and simulators, including demonstrations on matrices with sizes up to $1024\times1024$ in superconducting platforms~\cite{bravo2023variational}. Beyond proof-of-principle linear algebra, VQLS has also been proposed and applied in domain settings such as fluid dynamics~\cite{demirdjian2022variational,ye2024hybrid,bosco2024demonstration,meena2025assessing}, aligning closely with our focus on canonical flow problems on NISQ devices~\cite{meena2024solving,meena2024towards}. Despite this progress, there remains a gap between algorithmic descriptions and an end-to-end workflow that is robust enough for CFD-style matrices and practical enough to be embedded in hybrid quantum--HPC environments.

In particular, deploying VQLS at scale raises several interconnected questions that are not always addressed systematically in application-focused studies. First, the matrix $A$ must be encoded in a form suitable for the quantum subroutines used to evaluate the VQLS objective, and this step can become the dominant cost driver if it produces a large number of terms. Second, the choice of ansatz family impacts trainability and expressivity, and can interact strongly with problem structure: ansatz circuits that work for small, toy matrices may not generalize to structured CFD operators. Third, the performance and stability of the classical optimizer depend on the measurement noise model, the number of shots, and the cost landscape, which can vary substantially across matrix families and decomposition strategies. Finally, even when the quantum circuits themselves are relatively small, practical experimentation often requires running large ensembles of circuits across many optimization iterations, which naturally suggests HPC-style execution; however, the mapping between VQLS workloads and HPC parallelization strategies is subtle and can be counterintuitive.

%% file: Sections/Methodology.tex
\section{Methodology and Problem Setup} \label{sec:Methodology}
\begin{figure*}
  \centering
  \includegraphics[width=1\linewidth]{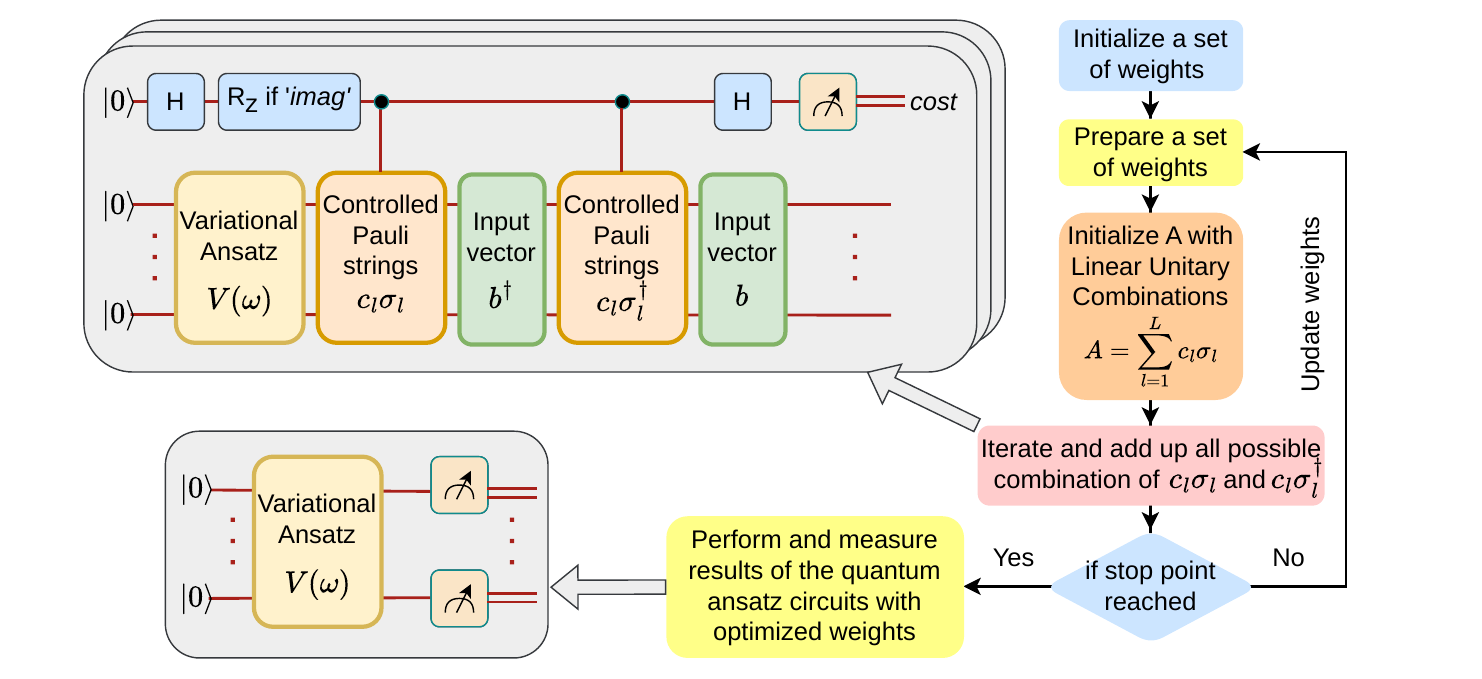}
  \caption{VQLS workflow combining quantum circuit components with classical optimization steps to
    approximate the solution to $A\ket{x} = \ket{b}$. The algorithm iterates the quantum circuits by updating the
    weights to minimize the overall cost. Once optimization converges, the measurement of the Variational
  ansatz reveals the solution state $\ket{x}$.}
  \label{fig:vqls_schematic}
\end{figure*}
\subsection{Algorithm Overview}
The objective of the VQLS algorithm is to find the state vector $\ket{x}$ for $x \in \mathbb{R}^{N}$, using a hybrid quantum--classical optimization approach. Given a guess for the vector, $\ket{\tilde{x}}$, a corresponding vector $\ket{\tilde{b}}$ can be found by
\begin{align}
  \ket{\tilde{b}} := \frac{A\ket{\tilde{x}}}{\sqrt{\bra{\tilde{x}}A^\dagger A\ket{\tilde{x}}}} \approx \ket{b}.
\end{align}
The goal is to optimize $\ket{\tilde{x}}$ to maximize the overlap between $\ket{b}$ and $\ket{\tilde{b}}$, given by the objective function
\begin{align}
  C = 1 - \left| \braket{b | \tilde{b}} \right|^2.
\end{align}
The state $\ket{\tilde{x}}$ is represented by a variational quantum circuit $V$ using a finite number of classical real parameters $w$, given by
\begin{align}
  \ket{\tilde{x}} = V(w)\ket{0}.
\end{align}
The parameters $w$ are optimized using a classical approach to minimize $C$. Further details of the cost function and different routes to approach the problem setup are elaborated in the original VQLS work \cite{bravo2023variational}. The key components of the methodology that this work will focus on are: (1) the representation of the $A$ matrix, (2) the nature of the variational circuit or ansatz, and (3) the choice of the classical optimizer. The overall schematic of the VQLS algorithm is demonstrated in Figure~\ref{fig:vqls_schematic}.

\subsection{Matrix preparation using linear combination of unitaries}

For the VQLS algorithm to be applicable, the $A$ matrix (and $\ket{b}$) should be encoded efficiently. One approach is to represent or decompose $A$ into a linear combination of unitaries (LCU), like the combination of Pauli operators
\begin{align}
  A = \sum_{l=1}^Lc_l\sigma_l.
\end{align}
Given any $A$, in our implementation we find coefficients of all possible combinations of $\mathbb{I}$ and Pauli $X$, $Y$, and $Z$ operators for the given basis dimension of the system as the trace inner product of each operator combination with $A$, normalized by the number of combinations. As we will see for the fluid flow problem, this creates challenges.

The number of LCU terms increases as the dimension of the matrix increases. In the worst-case scenario, a randomized dense square matrix with dimension $2^k$ requires $4^k$ LCU terms after decomposition, leading to inefficiency and poor scalability. Moreover, finding the Pauli terms and their coefficients is computationally heavy if not optimized. In the brute-force approach, one iterates over the full Pauli-string basis to calculate all coefficients, then drops the terms whose magnitude is below a threshold. The advantage of this method is that the input only needs to be a square matrix of dimension $2^k\times 2^k$. The detailed algorithm pseudocode is shown in Alg.~\ref{alg:naive_LCU}. Despite being easy to implement and requiring no special properties of the input matrix (Hermitian or unitary), this approach enumerates all Pauli strings and thus incurs excessive memory and computation overhead.

We adopt a memory- and compute-efficient approach proposed by Georges et al.~\cite{georges2025paulidecomp}, where the LCU decomposition only needs enough memory to store the input matrix and the computation of the Pauli terms uses fixed auxiliary memory, independent of matrix size. The improved algorithm leverages the Walsh--Hadamard transform and requires the input matrix to be Hermitian ($A=A^\dagger$). The basic idea is to use the Fast Walsh--Hadamard Transform (FWHT). The Pauli terms $\{I,X,Y,Z\}$ are indexed by two binary indices $r, s$ using $P(r_j, s_j) = i^{r_j s_j}X^{r_j}Z^{s_j}$. With this indexing, $(r_j,s_j)=(0,0),(1,0),(0,1),(1,1)$ correspond respectively to $I,X,Z,Y$ on qubit $j$. Then, the coefficients are calculated by:

\begin{align}
  \alpha_{r,s} &= \frac{i^{-|r \wedge s|}}{2^n} \sum_{q=0}^{2^n-1} a_{q \oplus r, q} (H^{\otimes n})_{q,s}. \label{eq:pauli_coeff_explicit}
\end{align}
where $a_{q \oplus r, q}$ denotes the XOR column permutation of the element $a_{r, q}$ in the input matrix $A$, and $H$ is the standard Hadamard gate with matrix expression
$H =
\begin{bmatrix}
  \begin{smallmatrix}1 & 1 \\\\1 & -1
  \end{smallmatrix}
\end{bmatrix}$.

To further improve scalability, we implemented shared memory for multiple workers so that the system stores the input matrix once while all cores compute Pauli terms and coefficients in an embarrassingly parallel manner.

\begin{algorithm}[bt!]
  \caption{Naive Pauli decomposition algorithm.}
  \label{alg:naive_LCU}
  \begin{algorithmic}[1]
    \Require Square matrix $A$, tolerance $\epsilon$.
    \Ensure List of Pauli terms $P$ and coefficients $C$.

    \State $N \gets \text{dimension}(A)$
    \State $n \gets \lceil \log_2(N) \rceil$
    \If {$N \neq 2^n$}
    \State Pad $A$ with zeros to size $2^n \times 2^n$.
    \EndIf

    \State Empty array of P, C
    \For {each $\sigma \in \{I, X, Y, Z\}^{\otimes n}$}
    \State $P_{\text{curr}} \gets \sigma_1 \otimes \sigma_2 \otimes \dots \otimes \sigma_n$
    \State $c \gets \frac{1}{2^n} \mathrm{Tr}(P_{\text{curr}} \cdot A)$

    \If {$|c| \ge \epsilon$}
    \State Append $c$ to $C$
    \State Append $P_{\text{curr}}$ to $P$
    \EndIf
    \EndFor
    \State \Return $P, C$
  \end{algorithmic}
\end{algorithm}

By adopting FWHT, the computational complexity is reduced to $\mathcal{O}(n^2\log n)$ with fixed auxiliary memory for the decomposition. This algorithm scales better than the naive approach and reduces the matrix-initialization bottleneck that often dominates practical QLSA workflows. We evaluate its performance in Section~\ref{sec:results}. The detailed algorithm is shown in Algorithm~\ref{alg:pauli_FWHT}.

\begin{algorithm}[bt!]
  \caption{Fast Walsh–Hadamard transform-based Pauli Decomposition}
  \label{alg:pauli_FWHT}
  \begin{algorithmic}[1]
    \Require $A=(a_{p,q})\in\mathbb{C}^{2^n\times 2^n}$
    \Ensure Pauli coefficients $\alpha_{r,s}$ (stored in-place)

    \For{$q = 0$ to $2^n-1$}
    \State \textbf{XOR permutation:}
    \For{$r = 0$ to $2^n-1$}
    \State $a_{r,q} \leftarrow a_{r\oplus q,\,q}$
    \EndFor
    \EndFor

    \For{$r = 0$ to $2^n-1$}
    \State \textbf{Walsh--Hadamard transform on row $r$:}
    \State $a_{r,*} \leftarrow H^{\otimes n} a_{r,*}$
    \EndFor

    \For{$r,s = 0$ to $2^n-1$}
    \State \textbf{Phase and normalization:}
    \State $a_{r,s} \leftarrow \dfrac{i^{-|r\wedge s|}}{2^n}\, a_{r,s}$
    \EndFor

  \end{algorithmic}
\end{algorithm}

\subsection{Minimization of decomposition terms}
Even though the Pauli strings and coefficients can be generated efficiently through the improved algorithm, the number of Pauli terms increases the total number of circuits executed on the quantum computer, leading to inefficiency in VQLS. Therefore, a decomposition method with fewer terms is desirable for better scalability. In this paper, we use a decomposition that requires at most two terms for an arbitrary square matrix of size $2^n\times 2^n$. This decomposition is based on the singular value decomposition (SVD) and expresses $A$ as the sum of two unitaries with equal weights, as derived below.

For an arbitrary square real matrix $A$, we apply a single SVD,
\begin{align}
  A = U \Sigma V^T,
\end{align}
where $U$ and $V$ are orthogonal (and hence unitary over the reals), and $\Sigma$ is a diagonal matrix of non-negative singular values $\sigma_j$. To map these singular values onto the complex unit circle, we first normalize by the spectral norm $\|A\|_2$,
\begin{align}
  \tilde{\Sigma} = \frac{\Sigma}{\|A\|_2},
\end{align}
so that all entries of $\tilde{\Sigma}$ are bounded by 1. We then construct two diagonal matrices by introducing an imaginary phase,
\begin{align}
  D_{\pm} = \tilde{\Sigma} \pm i \sqrt{I - \tilde{\Sigma}^2}.
\end{align}
Since $\tilde{\sigma}_j^2 + (1-\tilde{\sigma}_j^2) = 1$, every diagonal element of $D_{\pm}$ has magnitude 1, so $D_+$ and $D_-$ are unitary. Defining $U_{\pm} = U D_{\pm} V^T$, each $U_{\pm}$ is unitary as a product of three unitaries. Summing cancels the imaginary part,
\begin{align}
  U_+ + U_- = U(2\tilde{\Sigma})V^T = \frac{2A}{\|A\|_2},
\end{align}
and isolating $A$ yields the exact two-term LCU,
\begin{align}
  A = \frac{\|A\|_2}{2} U_+ + \frac{\|A\|_2}{2} U_-.
\end{align}
With this decomposition, we only need one extra preparation qubit along with the encoding qubits of the two unitaries for the computation and lead to the minimal number of LCU terms. A thorough analysis is described in Section~\ref{sec:results}.


\subsection{Quantum Ansatz Selection}

\begin{figure*}
  \centering
  \includegraphics[width=\linewidth]{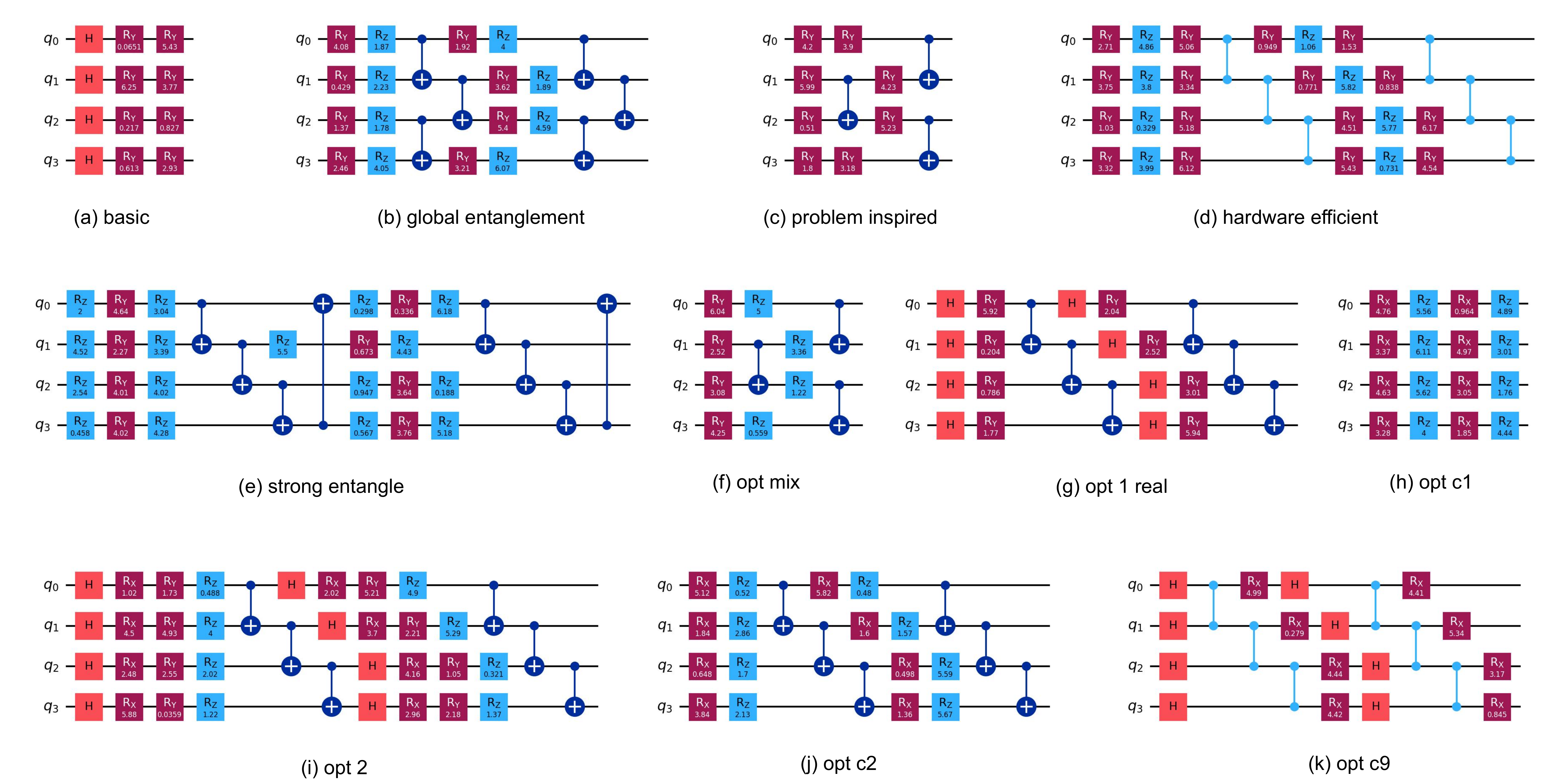}
  \caption{Different quantum ansatz setups with the corresponding expressibility. All quantum ansatz shown contain 4 qubits with 2 layers each. The numbers in the quantum gates are the parameters of the quantum ansatz.}
  \label{fig:ansatz_all}
\end{figure*}

The form of the particular ansatz circuit is also the key to capturing the essential physics of the state vector. There have been several works in the literature providing guidance on the choice of the ansatz for variational circuits. The work by \cite{sim2019expressibility} elaborates the choice based on quantitative metrics for expressibility and entangling capability of the ansatz. For the current implementation, we use amplitude encoding for representing $\ket{b}$.

In this work, we analyze several quantum ansatz families with different expressibility to study the trade-offs between trainability and expressibility. To quantify the ability of a parameterized quantum circuit (PQC) to uniformly explore the available Hilbert space, we compute its expressibility ($E$) using the statistical framework proposed by Sim et al.~\cite{sim2019expressibility}. This metric measures the deviation of the ansatz’s sampling distribution from the Haar-uniform distribution over pure quantum states in a $2^n$-dimensional Hilbert space. Specifically, we generate an empirical probability distribution
$\hat{P}_{\mathrm{PQC}}(F)$ of fidelities
\begin{equation}
  F = \left| \langle \psi(\boldsymbol{\theta}) \mid \psi(\boldsymbol{\phi}) \rangle \right|^2,
\end{equation}
obtained by sampling pairs of parameterized states using independently and
uniformly drawn parameters $\boldsymbol{\theta}$ and $\boldsymbol{\phi}$. This
empirical distribution is compared against the analytical fidelity distribution
for Haar-random pure states,
\begin{equation}
  P_{\mathrm{Haar}}(F) = (D-1)(1-F)^{D-2},
\end{equation}
where $D = 2^n$ is the Hilbert space dimension. The expressibility is defined as
the Kullback--Leibler (KL) divergence between the binned empirical distribution
and the corresponding Haar distribution,
\begin{equation}
  E = D_{\mathrm{KL}}\!\left(\hat{P}_{\mathrm{PQC}} \parallel P_{\mathrm{Haar}}\right)
  = \sum_i \hat{P}_{\mathrm{PQC}}(F_i)
  \ln\!\left(
    \frac{\hat{P}_{\mathrm{PQC}}(F_i)}{P_{\mathrm{Haar}}(F_i)}
  \right)
\end{equation}
A lower value of $E$ indicates that the circuit-generated states more closely approximate uniform sampling of the Hilbert space and therefore exhibit higher expressibility. Additionally, we define a normalized relative expressibility score by rescaling $E$ with respect to the Hilbert space dimension and binning resolution, such that higher values correspond to greater expressibility. Figure~\ref{fig:ansatz_all} provides several common quantum ansatz that we analyze in this work.

Another critical metric for a quantum ansatz is its entanglement capability, which quantifies the circuit's ability to generate entangled states. This value is determined by sampling a large set of circuit parameters from a uniform distribution and calculating the mean Meyer-Wallach measure across the resulting states \cite{sim2019expressibility}. A higher entanglement capability indicates superior performance, enabling the ansatz to access a broader range of quantum states during computation.

\subsection{Coherent Variational Quantum Linear Solver (cVQLS) and Block encoding}
\begin{figure}
  \centering
  \includegraphics[width=\linewidth]{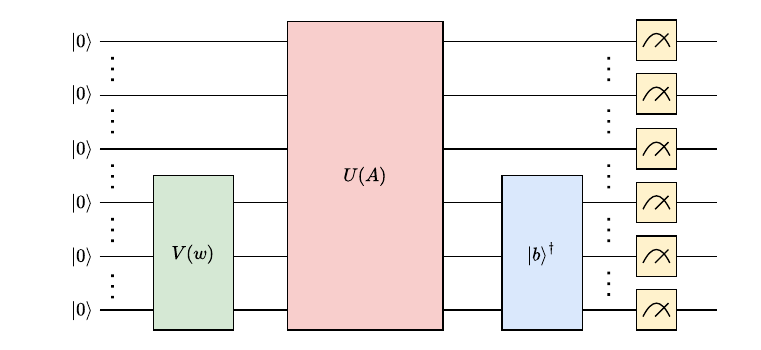}
  \caption{ Schematic representation of the Variational Quantum Linear Solver (VQLS) utilizing a Block Encoding strategy. The circuit replaces the standard Linear Combination of Unitaries (LCU) decomposition with a block-encoded unitary $U_A$ acting on the ancillary and system registers, coupled with the variational ansatz $V(\boldsymbol{\theta})$.}
  \label{fig:cvqls}
\end{figure}

\begin{figure}
  \centering
  \includegraphics[width=1\linewidth]{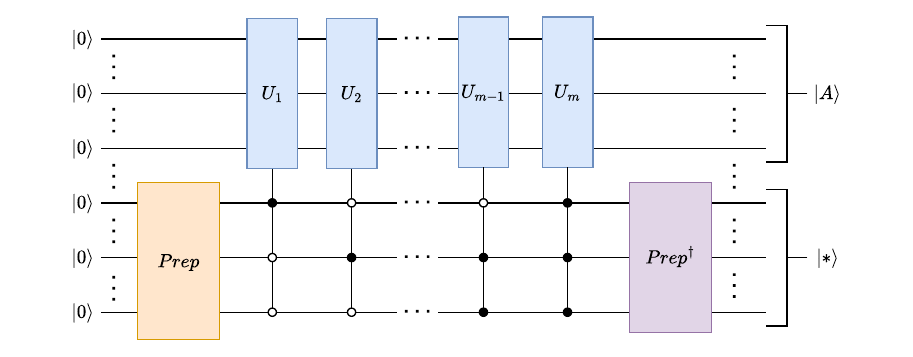}
  \caption{Circuit diagram for block encoding via LCU decomposition. The process consists of three stages: the state preparation unitary $Prep$ (acting on the ancilla), the selection operator $U_1, U_2 ... U_{m-1}, U_m$ (controlled by the ancilla), and the adjoint state preparation $Prep^\dagger$. The qubits that encode the selection operator contain the encoded original matrix $A$.}
  \label{fig:lcu_block_encoding}
\end{figure}

Another version of the quantum linear solver is to encode matrix $A$ into a single quantum circuit, which generates a larger circuit containing the encoded matrix $A$, the quantum ansatz, and the prepared state vector $\ket{b}$ \cite{mari2019cvqls}. Similar to VQLS, CVQLS prepares the quantum ansatz $V(w)$ such that $V(w)\cdot \ket{0} =\ket{x}$ and minimizes $1- U(A)\ket{V(w)} \cdot \ket{b}^\dagger$ by changing parameters $w$. The vector $\ket{x}$ can then be reconstructed by computing $V(w_{\min})$. The circuit diagram is shown in Figure~\ref{fig:cvqls}. To encode matrix $A$ into a single quantum circuit, it requires a block-encoding technique to initialize a larger unitary matrix $
\begin{smallmatrix}A =
  \begin{bmatrix}     A\ * \\    *\ *
  \end{bmatrix}
\end{smallmatrix}$, where the normalized matrix $A$ is a block of a larger matrix; this technique is called block encoding\cite{gilyen2019quantum, chakraborty2018power, camps2024explicit, sunderhauf2024block}. There exist several approaches to perform block encoding. The most straightforward approach is the built-in function of the Pennylane package, where it encodes
\begin{align}    U(A) &=
  \begin{bmatrix}  A & \sqrt{I-AA^\dagger} \\ \sqrt{I-A^\dagger A} & -A^\dagger
  \end{bmatrix}.
\end{align}
directly by assigning the input matrix and the corresponding qubits. We also implement block encoding through LCU decomposition. This approach realizes the target operator by interleaving a Select oracle between a Prepare oracle and its adjoint. The schematic is demonstrated in Fig.~\ref{fig:lcu_block_encoding}. However, when physically implement this algorithm, the Pennylane software package does not provide gradient calculation for such encoding method, and the encoding method requires similar resource to the block encoding of the SVD decomposition, where the SVD decomposition-based block encoding requires as well. Therefore, we will not include this method in Section \ref{sec:results} \cite{d2025towards}. 

\subsection{Experimental setup}
We implemented the VQLS algorithm with a single computing node on the Frontier supercomputer using PennyLane~\cite{bergholm2018pennylane} (version 0.38.0) with a PyTorch backend (Torch version 2.5.0) for gradient-based optimization. We use the SciPy package (version 1.15.3) for Scipy-based optimization. We used the HPC-enabled quantum simulator {\tt lightning.kokkos} to gain acceleration on Frontier HPC system.

In the current work, we evaluate the implementation using three different problems based on their complexity: (1) a system comprised of LCU for $A$, similar to that evaluated in \cite{bravo2023variational}, (2) a tridiagonal Toeplitz system, and (3) a canonical fluid flow problem. Following the analysis in \cite{meena2024solving}, we choose the fluid flow problem as the two-dimensional Hele--Shaw flow, which is an inviscid steady flow resulting in the simplification of the Navier--Stokes equations to a system of coupled linear partial differential equations.


%% file: Sections/Results.tex
\section{Results}\label{sec:results}
We evaluate the LCU decomposition routine and several variants of the VQLS algorithm introduced in Section~\ref{sec:Methodology}. To enable a fair comparison, we report wall-clock time, average memory usage, and peak memory usage for each method. For cVQLS, we use two compute nodes to assess scalability with the {\tt Lightning.kokkos} simulator.

\subsection{LCU decomposition} \label{sec:LCU_rst}
\begin{figure*}
    \centering
    \includegraphics[width=0.45\linewidth]{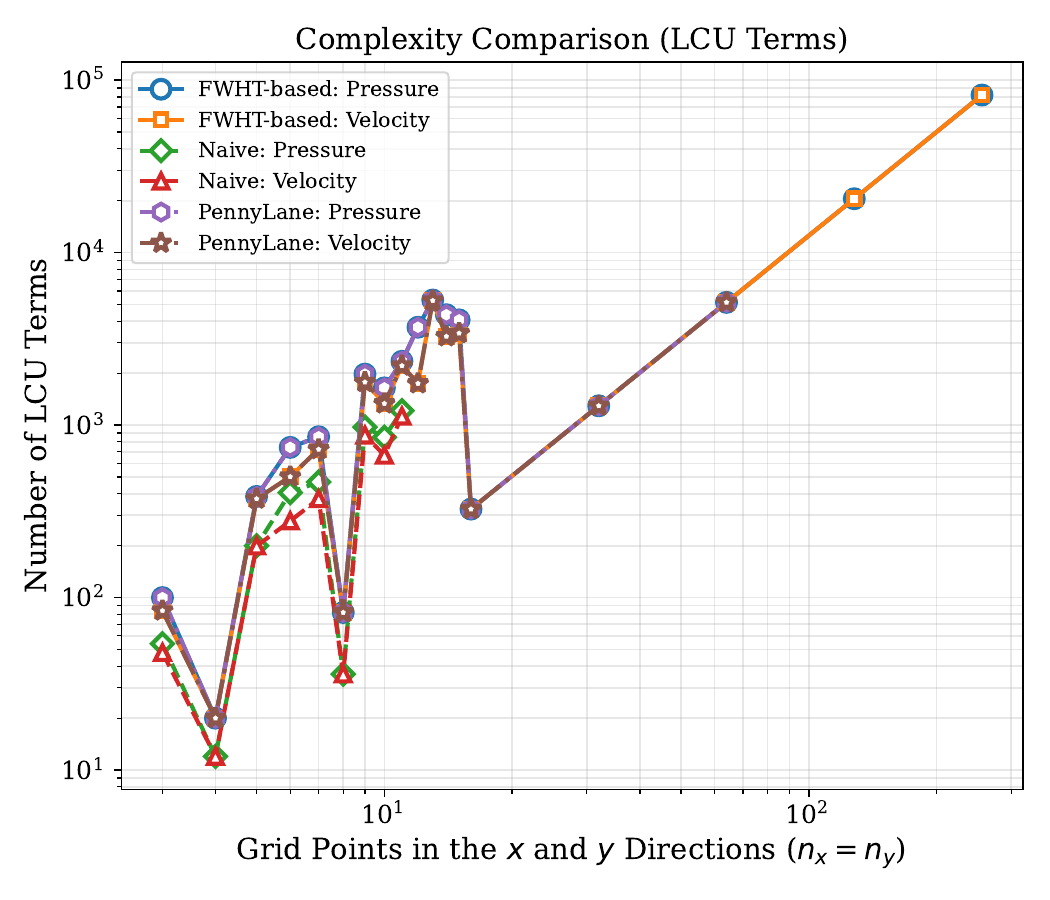}
    \includegraphics[width=0.45\linewidth]{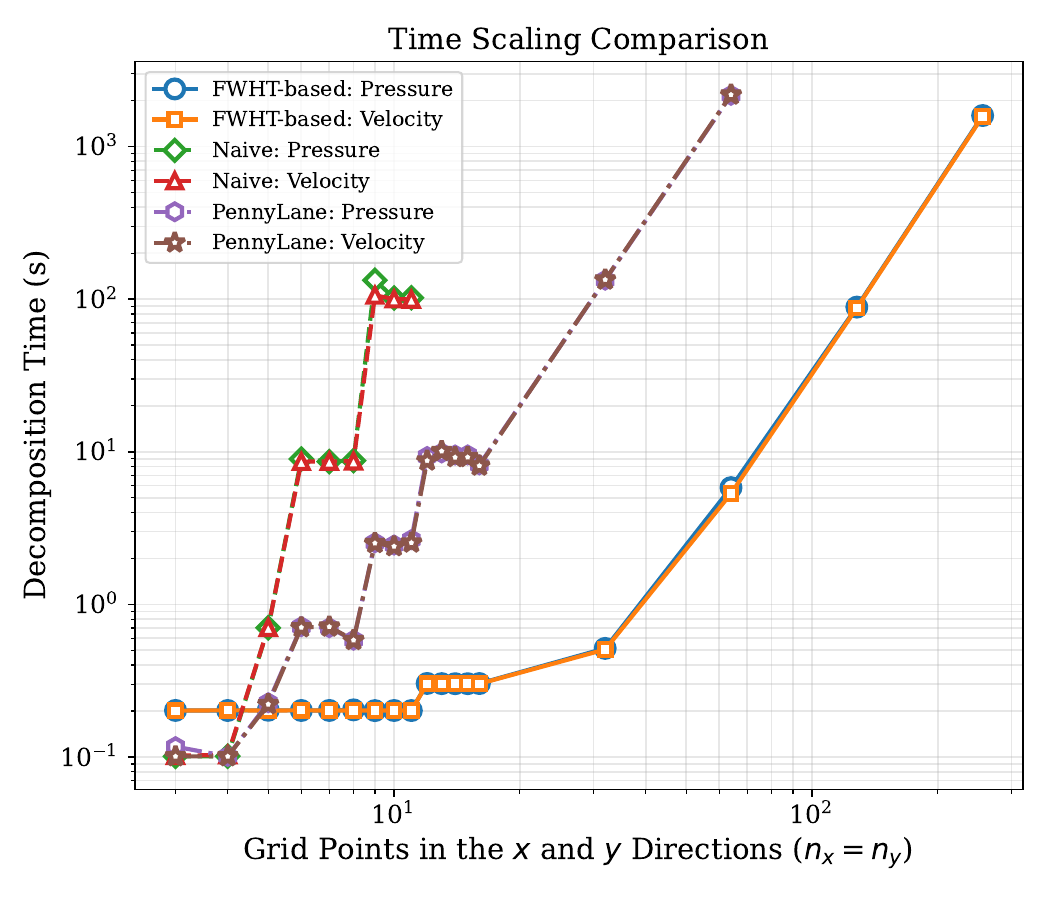}
    \includegraphics[width=0.45\linewidth]{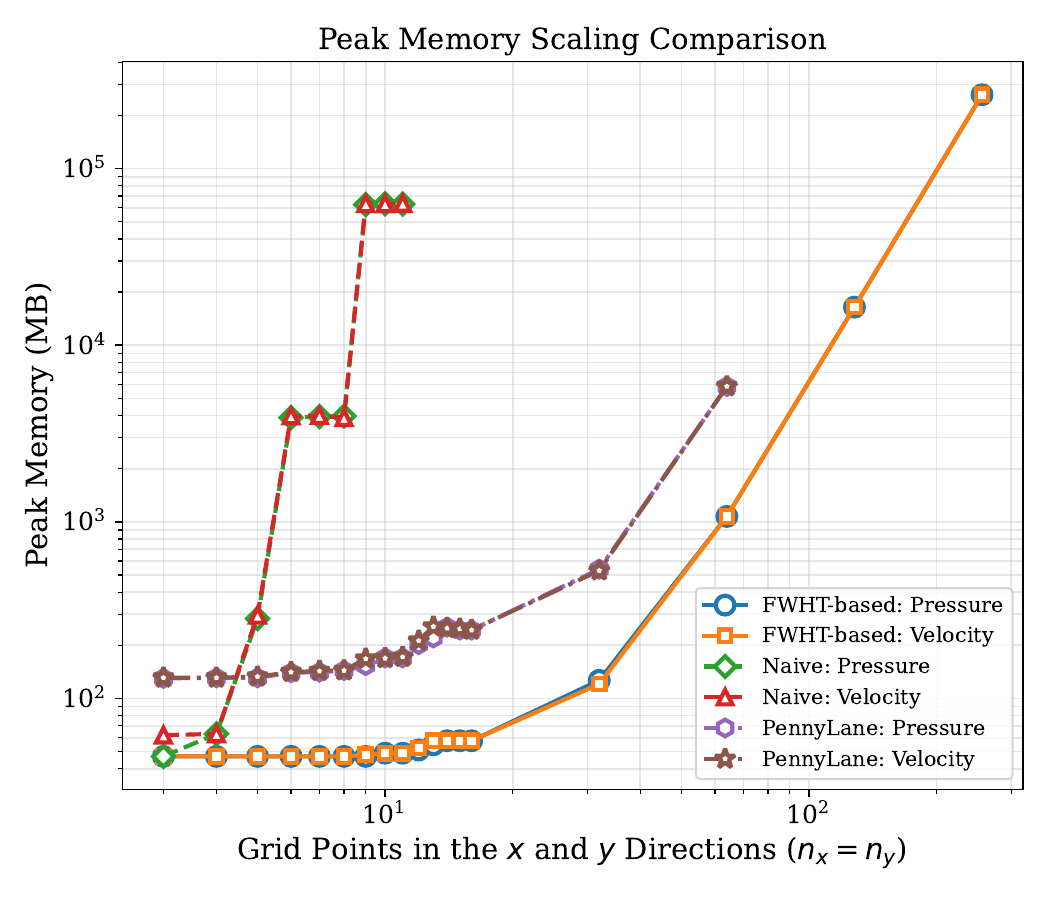}
    \includegraphics[width=0.45\linewidth]{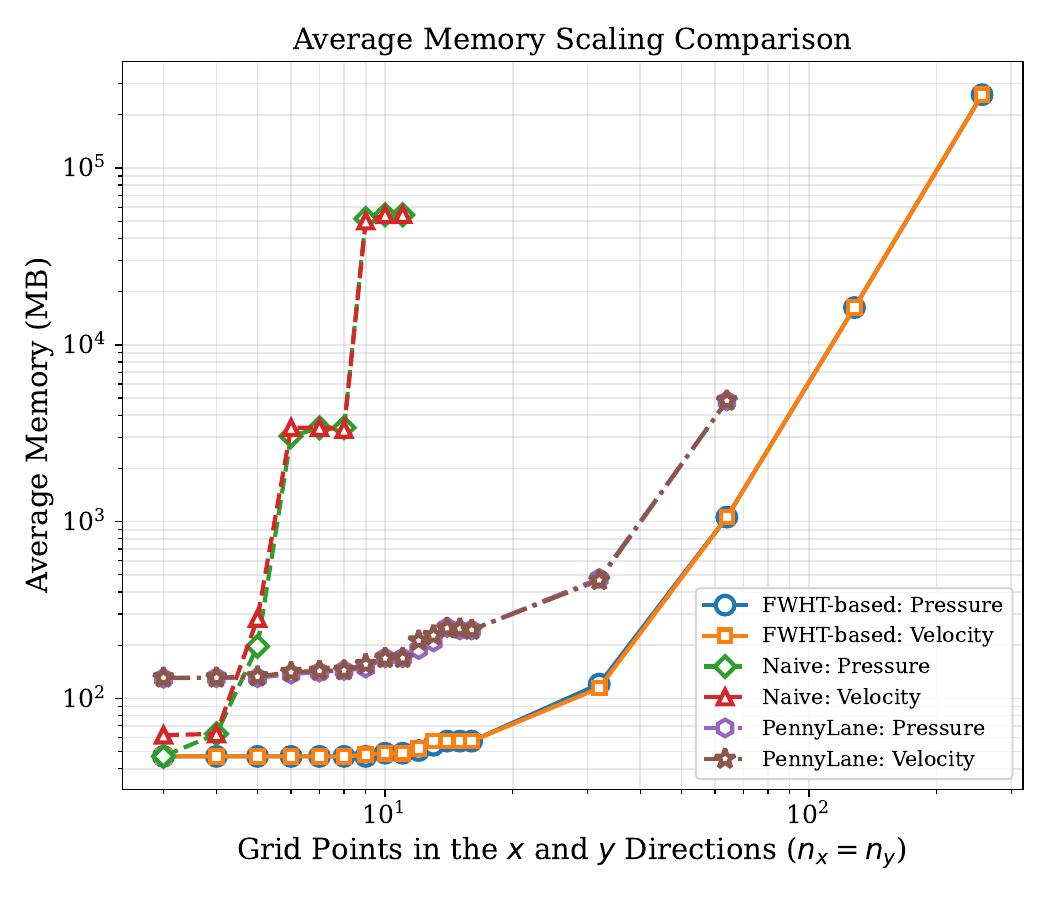}
    \caption{Pauli decomposition performance for a canonical Hele--Shaw Jacobian using the naive approach, the PennyLane-integrated approach, and the FWHT approach.}
    \label{fig:lcu_decomposition}
\end{figure*}

We first benchmark three LCU decomposition strategies: (1) a naive construction, (2) the PennyLane-integrated LCU decomposition, and (3) a Fast Walsh--Hadamard transform (FWHT) approach with CPU-core parallelization.
Figure~\ref{fig:lcu_decomposition} summarizes results for tridiagonal Toeplitz matrices and a canonical Hele--Shaw Jacobian with increasing grid dimensions $n_x$ and $n_y$. The naive approach supports up to $n_x=n_y=11$ and consistently yields the fewest Pauli terms, while the other methods typically require slightly more terms. In addition, when the grid dimension is a power of two ($n_x=n_y=2^k$ for $k=2,3,4,\dots$), the number of Pauli terms is minimized and scales approximately linearly with grid resolution; non-power-of-two sizes require padding to embed the matrix into a circuit, increasing the term count. 

In terms of runtime, the naive and PennyLane-integrated methods are comparable for small problems ($n_x=n_y\in\{3,4\}$) and are faster than the multi-threaded FWHT approach. As the problem size increases, however, the naive method fails to scale beyond $n_x=n_y=11$, making it the least scalable option. The PennyLane-integrated method scales better, but requires more than 2000~s to generate a case with $n_x=n_y=64$ and does not complete $n_x=n_y=128$ within a 2~h time limit. The FWHT-based decomposition is the most scalable approach, requiring less than 2000~s for $n_x=n_y=256$.
For both peak and average memory, the FWHT-based method outperforms the other approaches across all tested cases, achieving up to a $1298\times$ reduction at $n_x=n_y=11$ relative to the naive method. The PennyLane-integrated approach uses approximately $5\times$ more memory than FWHT across problem sizes. Across these metrics, the FWHT method provides the best overall scalability, while the naive and PennyLane-integrated methods can be competitive only for the smallest grid sizes.

\begin{figure}
    \centering
    \includegraphics[width=\linewidth]{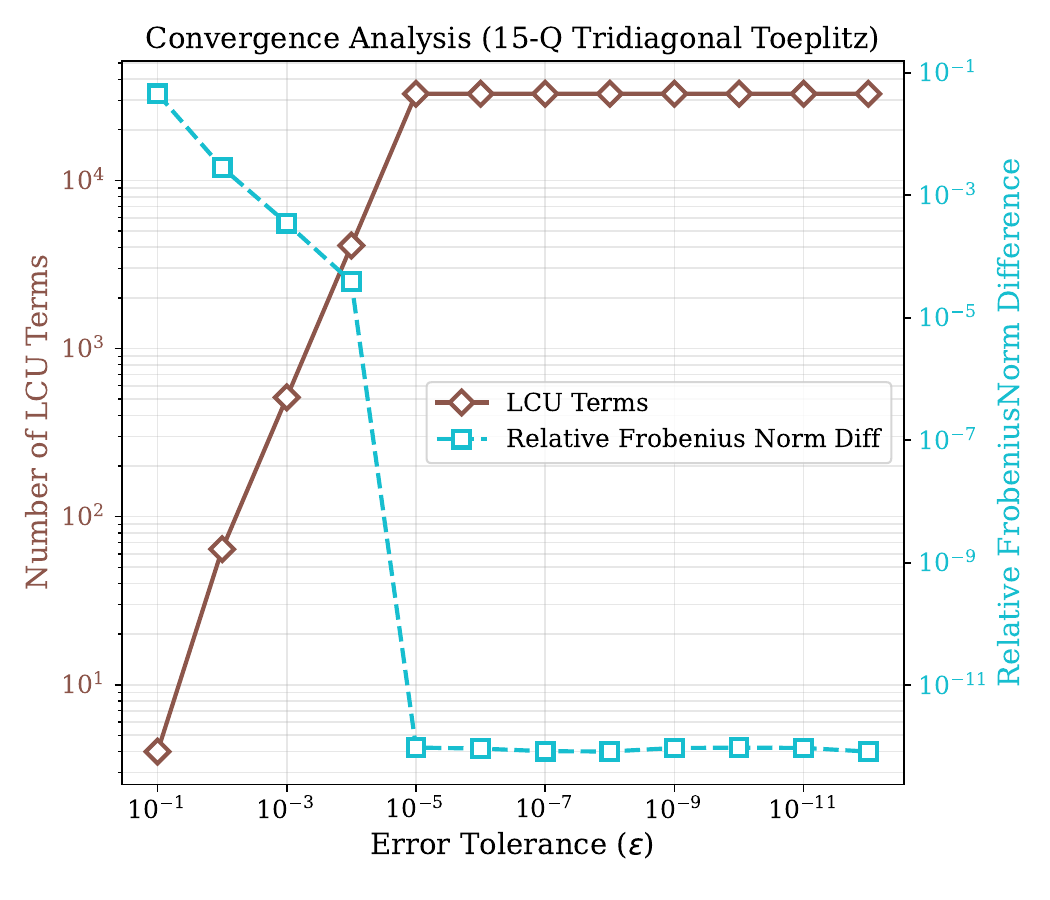}
    \caption{Number of Pauli LCU terms versus relative Frobenius-norm error.}
    \label{fig:LCU-approx}
\end{figure}

The proposed parallelized Pauli decomposition also supports an approximate mode, enabling fewer LCU terms and faster decomposition. We apply this approximate decomposition to a tridiagonal Toeplitz matrix of size $2^{15}\times 2^{15}$ and sweep the truncation tolerance from $10^{-12}$ to $10^{-2}$. Figure~\ref{fig:LCU-approx} reports the resulting runtime, memory usage, and approximation error measured by the relative Frobenius norm
$$
\frac{\lVert A-B\rVert_F}{\lVert A\rVert_F}.
$$
The trade-off between fidelity and term count is controlled by discarding coefficients whose magnitude falls below the tolerance threshold. For a tolerance of $10^{-1}$, the decomposition requires only four Pauli terms and achieves a relative Frobenius error of 0.045, indicating substantial compression with limited loss in fidelity. As the tolerance decreases from $10^{-2}$ to $10^{-4}$, the number of Pauli terms increases from 64 to 512 and 4096, and saturates at $10^{-5}$ with 32768 terms, while the relative Frobenius error drops below $10^{-11}$. Reducing the number of LCU terms directly decreases the number of circuits executed by VQLS, thereby reducing per-iteration cost and improving overall convergence time.

Figure~\ref{fig:LCU-approx} illustrates the accuracy--cost trade-off of approximate decomposition: increasing the tolerance reduces the number of retained Pauli terms and therefore reduces runtime and memory, at the expense of a larger relative Frobenius-norm error.


\subsection{Validation}
\begin{figure}[t]
 \centering
 \includegraphics[width=\columnwidth]{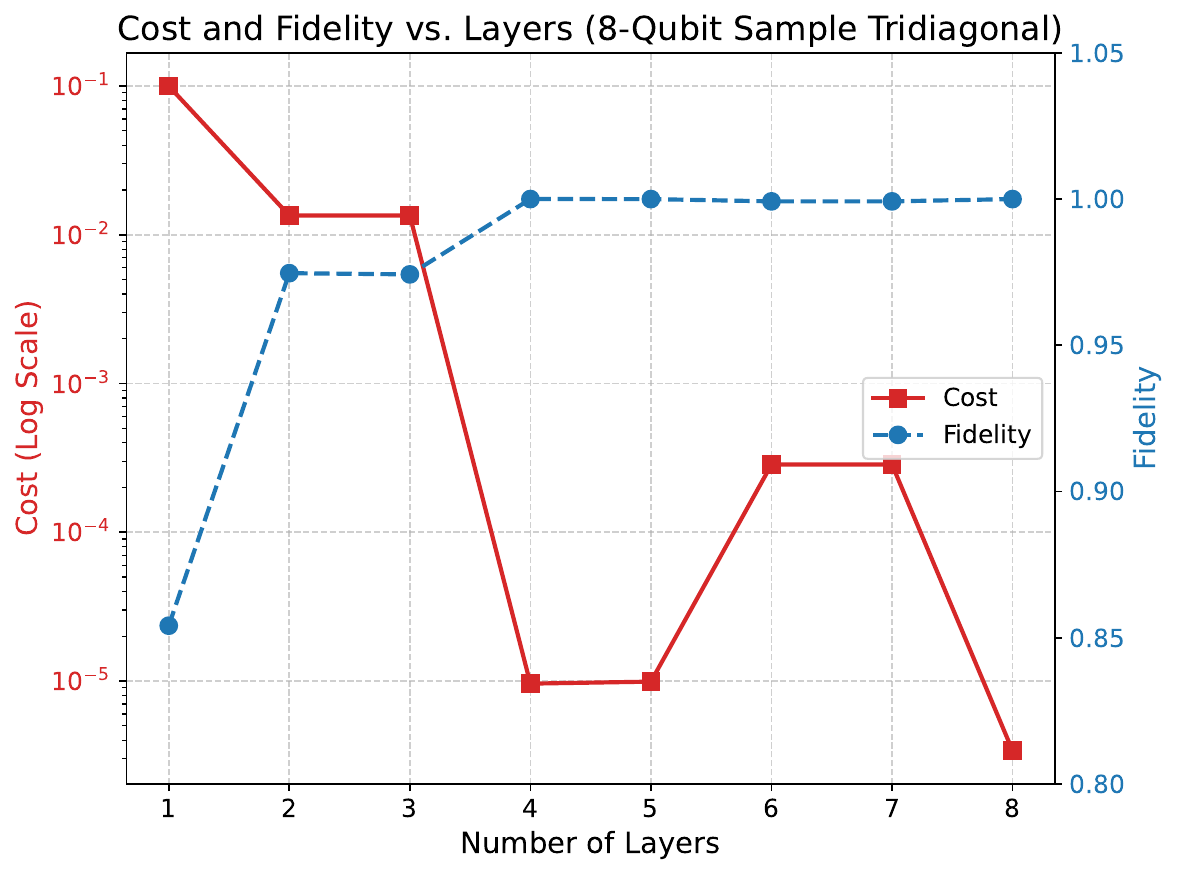}
 \caption{Effect of quantum ansatz layers on cost value and fidelity.}
 \label{fig:results-sample-fidelity}
\end{figure}

The validation of our LCU decomposition for a generalized $A$ matrix is computed  using the tridiagonal Toeplitz matrix with a system size of $2^8 \times 2^8$. We use the NAdam optimizer. We used different layers of the quantum ansatz `opt-mix' to perform the optimization to show the relationship between cost value and fidelity as shown in \fig\ref{fig:results-sample-fidelity}. The final cost value converges to $9.6\times 10^{-6}$ with a minimum of 4 ansatz layers, and the fidelity exceeds $0.9999$ with 4 or more layers, showing that the expressibility and convergence performance can increase by increasing the number of layers. The fidelity can reach to almost 1.00 when the cost value is below $10^{-3}$. However, after a certain number of layers, the increasing number of layers does not benefit to the final cost and fidelity result.

For the reduced LCU terms tridiagonal Toeplitz mentioned in the Section \ref{sec:LCU_rst} implemented for the VQLS algorithm, it shows similar convergence speed and reaches a cost of $1.5\times 10^{-6}$ with a fidelity of $0.99999$ relative to the classical solution, demonstrating negligible loss in accuracy while reducing computing overhead through the pruning strategy.


\subsection{Optimizer selection}
The choice of classical optimizer is crucial, particularly when navigating a complex loss landscape. Scipy-based methods such as COBYLA~\cite{powell1994direct,pellow2021comparison} can be attractive because they typically require only one function evaluation per iteration, but they may struggle on highly nonconvex objectives. Gradient-based methods such as stochastic gradient descent and NAdam often traverse such landscapes more effectively, but typically require many more circuit evaluations per optimization step to calculate the gradient process.

The implementation results in circuits with $n_b$ qubits, where $n_b$ is the number of qubits required to represent the state vector, $N = 2^{n_b}$, and the extra qubit is the ancilla. The depth of the circuit depends on the ansatz, and the number of circuit executions depends on the optimizer and how $A$ (and $\ket{b}$) are encoded. In our implementation of the LCU decomposition of $A$, for each function evaluation of the optimizer, there will be $2L^2 n_b$ circuit executions.

\begin{figure}[t]
 \centering
 \includegraphics[width=\columnwidth]{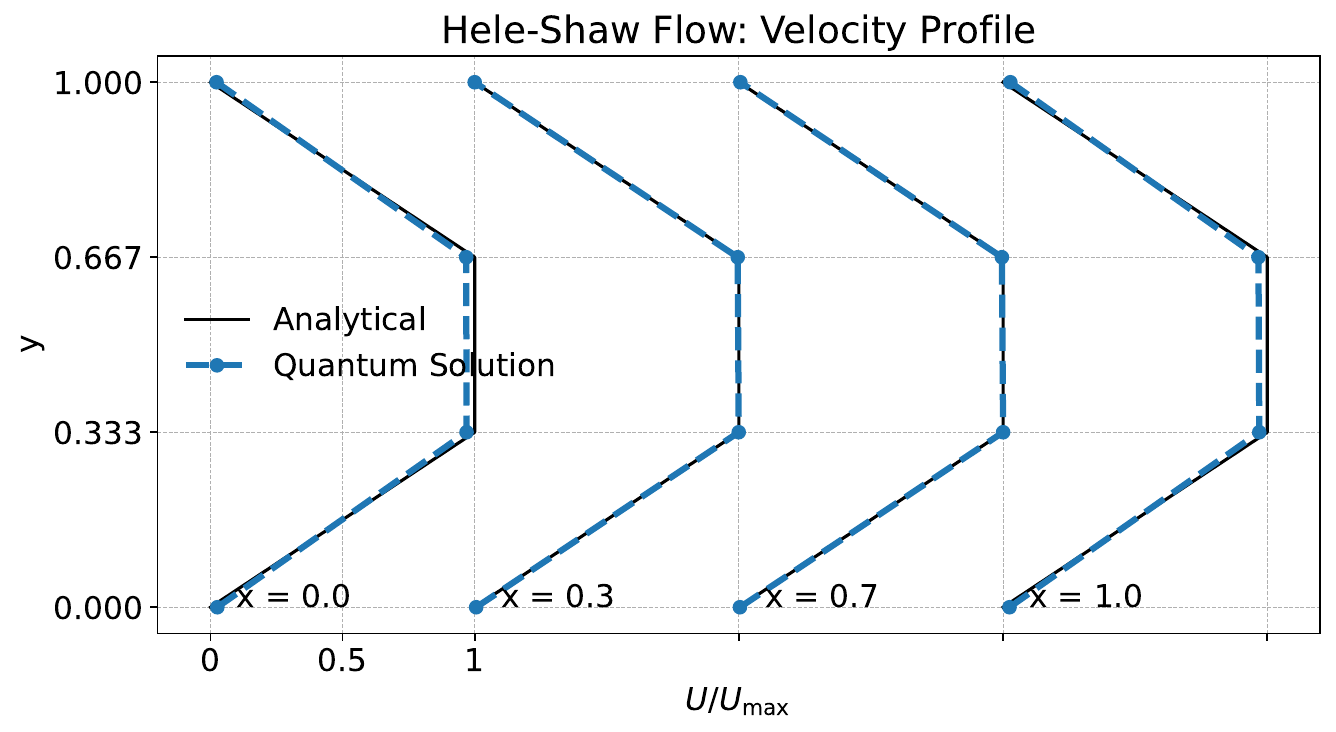}
 \caption{Comparing velocity reconstruction for the Hele--Shaw flow with $(n_x, n_y) = (4,4)$ grid points using VQLS algorithms.}
 \label{fig:results-HS}
\end{figure}

Given the validation of the implementation on simpler systems, we move on to solving the Hele--Shaw flow problem. Results of reconstructing the velocity field at distinct horizontal locations for a candidate flow setup with $(n_x,n_y)=(4,4)$ grid points are shown in \fig\ref{fig:results-HS}. The VQLS solution align with the classical solution well after the convergence with the final fidelity of 0.999438. 

\subsection{Quantum ansatz}
Selecting the right ansatz for fluid flow systems is challenging. Guidance can be deduced from the coefficients of the LCU decomposition of $A$. For both the LCU-based and tridiagonal Toeplitz systems, aside from having far fewer nonzero coefficients, the coefficients are purely real. Thus, the ansatz for these systems could benefit from gates operating predominantly on the real subspace. In contrast, the LCU decomposition of the Hele--Shaw $A$ yields both real and complex coefficients, potentially requiring higher expressibility to capture phase information.

Based on these deductions and guidance from \cite{sim2019expressibility}, we analyze 11 quantum ansatz families in terms of expressibility by calculating KL divergence, circuit statistics, and efficiency, as shown in Table~\ref{tab:kl_div_ansatz}. We define the efficiency metric as the KL divergence multiplied by the number of CX gates, since two-qubit gates take significantly longer and are typically less accurate than single-qubit gates. From the table, the `hardware\_efficient' ansatz has the lowest KL divergence, meaning that it can generate the broadest set of states among the listed ansatz while using only six CX gates, yielding the best efficiency. The `basic' and `opt-c1' ansatz have the worst expressibility due to the lack of CX gates to create entanglement.

\begin{table}[h!]
\centering
\caption{Comparison of Expressibility (KL Divergence) and Entanglement capability for various Quantum Ansatz ($N=4, L=2$).}
\label{tab:kl_div_ansatz}
\scalebox{0.9}{
\begin{tabular}{l c c}
        \toprule
        \textbf{Method} & \textbf{KL Divergence} & \textbf{Entanglement Capability} \\
        \midrule
        strong-entangle      & 0.0062 & 0.8474 \\
        opt-2                & 0.0167 & 0.7043 \\
        global\_entanglement & 0.0168 & 0.6850 \\
        hardware\_efficient  & 0.0204 & 0.6047 \\
        opt-c2               & 0.0225 & 0.6907 \\
        opt-c1               & 0.2101 & 0.0000 \\
        opt-mix              & 0.2722 & 0.4717 \\
        problem\_inspired    & 0.4293 & 0.3656 \\
        opt-c9               & 0.4318 & 1.0000 \\
        opt-real-1           & 0.5602 & 0.3748 \\
        basic                & 0.6309 & 0.0000 \\
        \bottomrule
    \end{tabular}
    }
\end{table}

\begin{figure}
    \centering
    \includegraphics[width=\columnwidth]{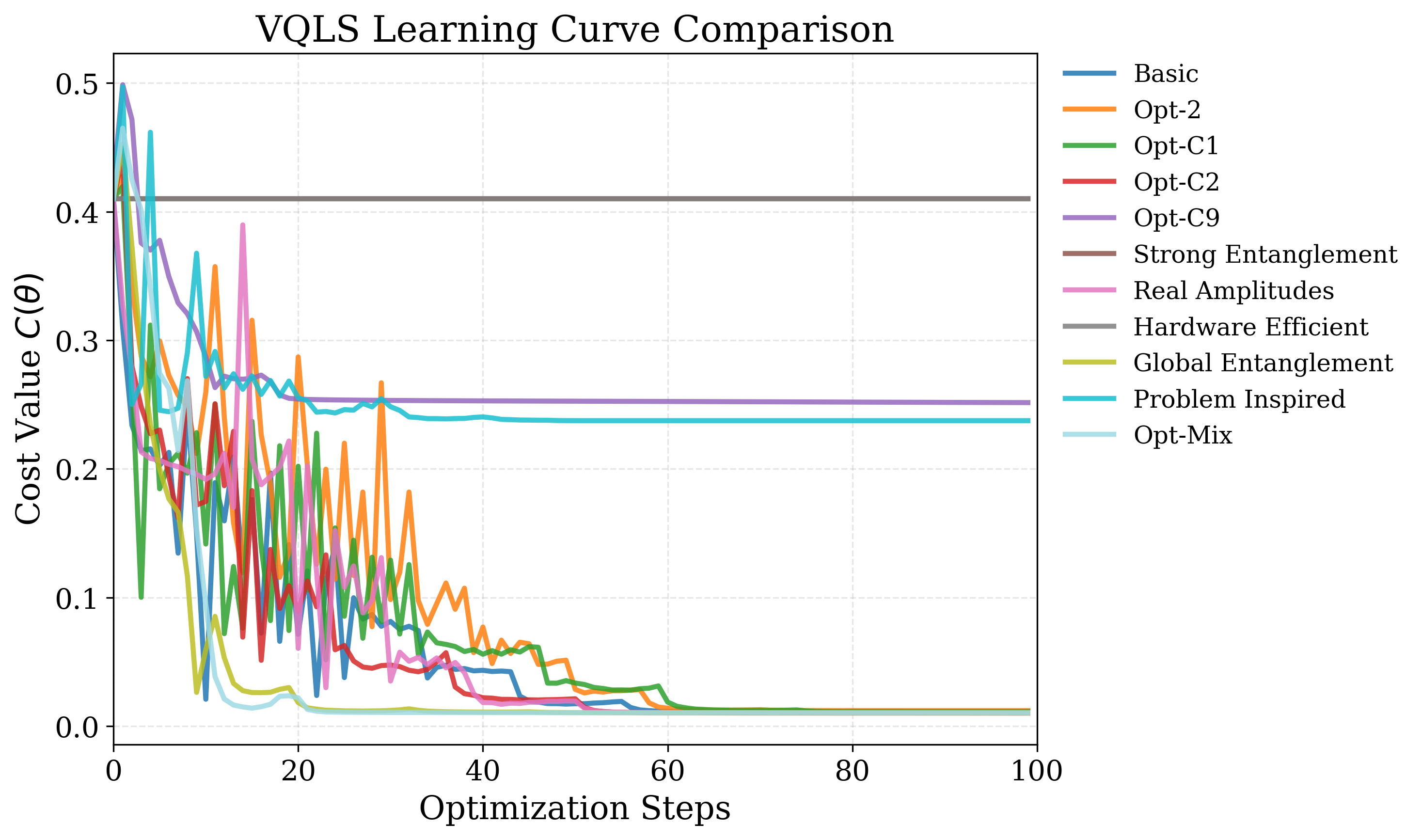}
    \caption{Torch-based optimizer (NAdam) learning curve comparison of all quantum ansatz mentioned in Figure~\ref{fig:ansatz_all} using the 2D Hele--Shaw problem with $n_x=n_y=4$.}
    \label{fig:learning_curve}
\end{figure}

Figure~\ref{fig:learning_curve} compares optimizer convergence across ansatz choices on the $n_x=n_y=4$ Hele--Shaw instance. We use it to identify ansatz that reliably reach low cost within the fixed iteration budget and to highlight cases that either converge rapidly (steeper initial decrease) or stagnate (plateau above the target threshold).

\begin{figure}[t]
 \centering
 \includegraphics[width=\columnwidth]{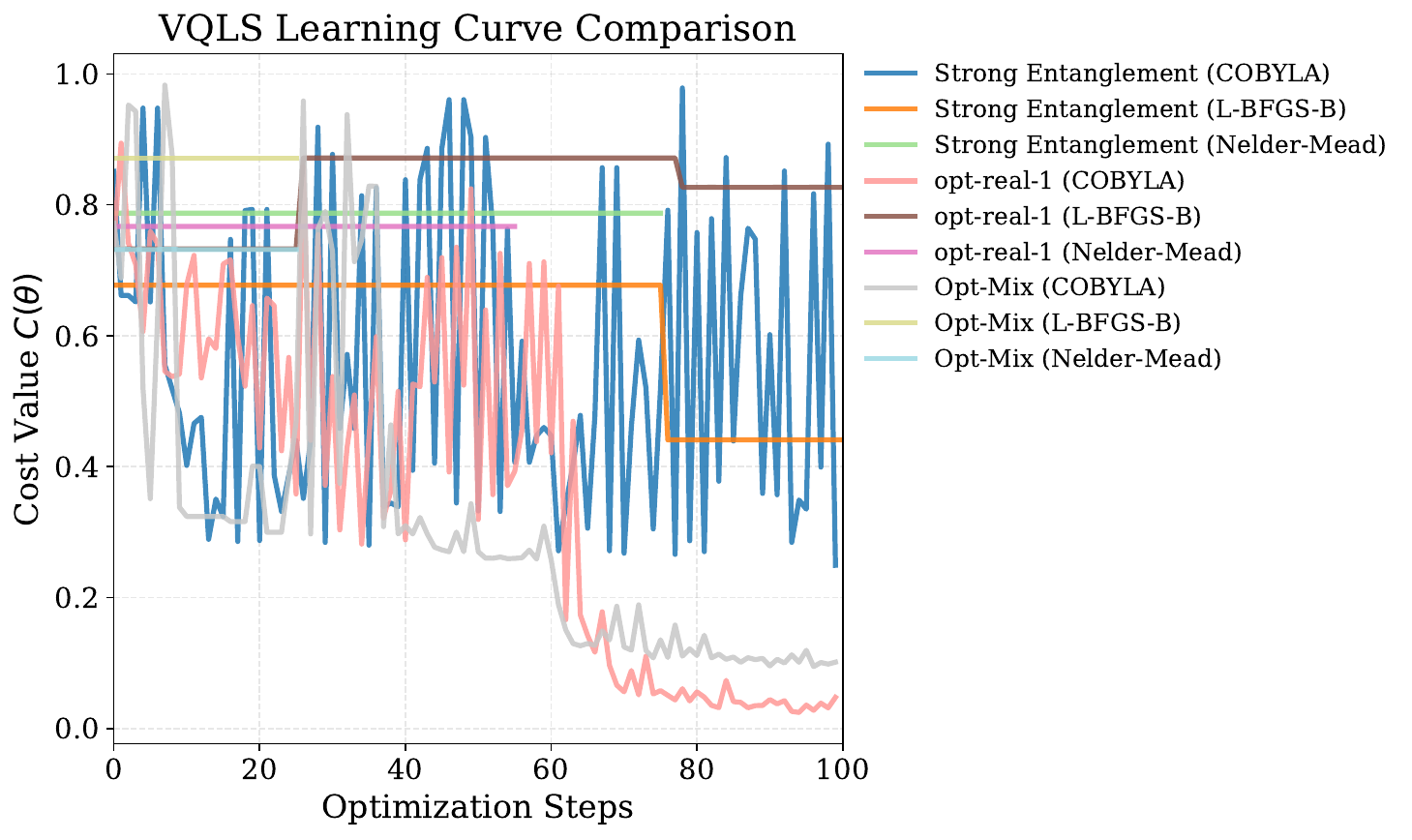}
 \caption{Learning curve of Scipy optimizer (COBYLA, L-BFGS-B, Nelder-Mead) performing minimization of quantum algorithms using three layers of quantum ansatz of `Strong Entanglement', `Real Amplitudes', and ` Opt-Mix'.}
 \label{fig:learning-curve-optimizer-gradient-free}
\end{figure}

We simulated a canonical Hele--Shaw flow problem on a $4\times4$ grid using the quantum ansatz listed in Table~\ref{tab:kl_div_ansatz}. Optimization was performed using the torch-based NAdam optimizer integrated within PyTorch, with learning curves illustrated in Figure~\ref{fig:learning_curve}. The majority of the ansatz converged within 100 optimization steps. Notably, Opt-Mix and Global Entanglement demonstrated the fastest performance, converging in fewer than 40 steps. The Basic, Opt-2, Opt-C1, and Opt-C2 ansatz required between 50 and 60 steps. Conversely, Problem Inspired, Opt-C9, and Strong Entanglement failed to reach a cost function value below 0.01 within the 100-step limit. Interestingly, convergence performance does not align with the KL divergence values observed in Table~\ref{tab:kl_div_ansatz}. For instance, while the Strong Entanglement ansatz exhibits high expressibility, it showed poor convergence for the Hele--Shaw problem. In contrast, the Opt-C1 and Basic ansatz converged successfully despite lacking entanglement.
 
We also performed Scipy-based optimizers `COBYLA', `L-BFGS-B', `Nelder-Mead', integrated in `Scipy' python package using three different quantum ansatz using three layers of`Strong Entanglement', `Real Amplitudes', and ` Opt-Mix' solving a $4$ by $4$ grid points on Figure \ref{fig:learning-curve-optimizer-gradient-free}. From the figure, we observe a inefficiency performing optimization compared to the torch-based approach, as they generally require more function evaluations or fail to reach the same low cost values within the iteration limit.

\subsection{HPC timing scalability of standard VQLS and cVQLS}
We performed simulations on the Frontier supercomputer. We used both {\tt lightning.qubit} and {\tt lightning.kokkos} to measure time per VQLS iteration and assess scalability with the available simulators. In this setting, Scipy-based optimizers such as COBYLA enable batching and parallel circuit evaluation, which benefits from CPU multi-threading to improve the performance on HPC system, whereas Torch-based optimization is effectively sequential and therefore less amenable to circuit-level parallelism.

\begin{figure}[t]
 \centering
 \includegraphics[width=0.46\textwidth]{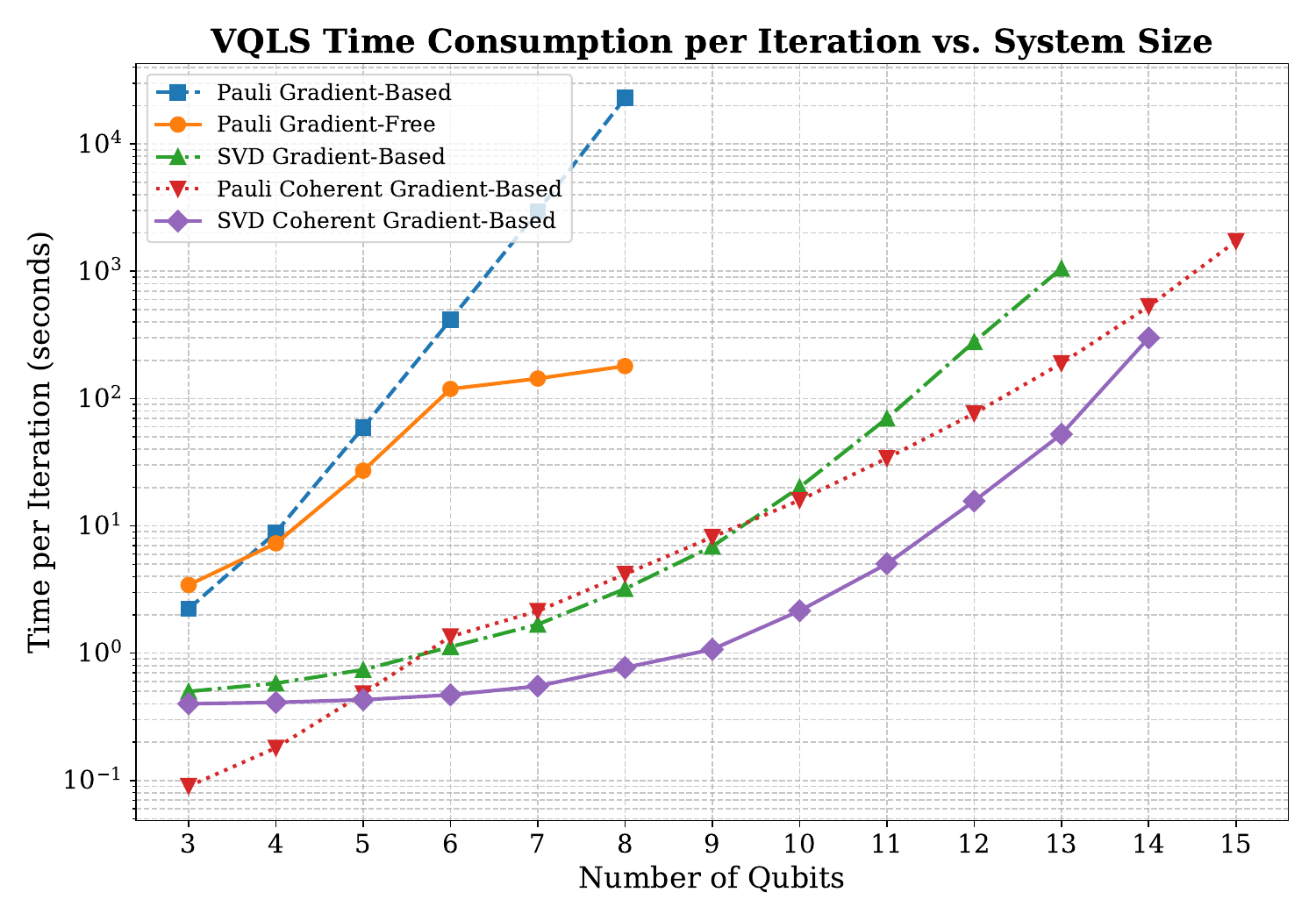}
 \caption{Time consumption scaling per iteration of standard and coherent VQLS with Torch-based and Scipy-free optimizers from 3 to 15 qubits, using both Pauli and SVD decomposition approaches.}
 \label{fig:time-vqls-all}
\end{figure}

Figure \ref{fig:time-vqls-all} demonstrates the performance of average time consumption per iteration of VQLS and coherent VQLS algorithm using the pennylane {\tt lightning.kokkos} and {\tt lightning.qubit} for Scipy-based VQLS algorithm to achieve the best performance with respect to time performance, extending the benchmark from 3 up to 15 qubits. We specifically implemented the pruned Pauli decomposition for the standard VQLS algorithm with the tolerance 0.01 to reduce the number of circuit execution pressure. From the Figure, the standard torch-based (NAdam) VQLS with Pauli decomposition requires the most time consumption and is the least scalable on time among all methods mentioned; its runtime grows so rapidly with system size that it becomes impractical well before reaching 15 qubits. The standard VQLS with Scipy-based optimizer (COBYLA) algorithm has better scaling performance compared with the torch-based algorithm.

After the implementation of the SVD-based decomposition, the number of circuits required per iteration is reduced dramatically compared with the Pauli decomposition, yielding better scalability and over a $10{,}000\times$ time reduction on an 8-qubit tridiagonal Toeplitz matrix, with the advantage widening further at larger system sizes up to 15 qubits. For the coherent VQLS that only requires one circuit to calculate the cost value, it has better timing performance compared with the standard VQLS with Pauli decomposition. However, the slope increases as the number of qubits increases because the number of qubits required for the state preparation and encoding circuits grows dramatically during execution, leading to inefficiency for large-qubit simulation. For the coherent VQLS using SVD-based block encoding, it has the best performance among all mentioned VQLS variants with the minimum time consumption requirement. However, the slope increases dramatically as the SVD decomposed unitary matrix encoding requires large amount of memory, leading to `out-of-memory' error for GPU simulation. The coherent VQLS with Pauli encoding achieved lowest slope among all approaches, and can simulate up to 15-qubits linear systems problem within the 2-hours wall time.

%% file: Sections/Conclusion.tex
\section{Concluding Remarks and Future Directions}\label{sec:Conclusion}

We presented an end-to-end study of the Variational Quantum Linear Solver (VQLS) for CFD-relevant linear systems on a hybrid quantum--HPC stack. The work was organized around the three challenges identified in the introduction---matrix encoding cost, ansatz selection, and HPC deployment---and we summarize our findings along the same axes.

\textbf{Matrix encoding.} On the tridiagonal Toeplitz system and the two-dimensional Hele--Shaw flow, the FWHT-based parallel Pauli decomposition reduces peak memory by up to $1298\times$ over the naive LCU on an $11\times 11$ grid while attaining $\mathcal{O}(n^2\log n)$ scaling. The SVD-based two-term LCU minimizes the number of LCU terms (two) and qubit count required for the encoded circuit, yielding more than $10{,}000\times$ per-iteration speedup over standard Pauli-based VQLS at 8 qubits and the best per-iteration timing among the four strategies we tested. Each approach has a distinct failure mode: Pauli-based encodings still produce many non-zero terms in the worst case (up to the number of matrix entries), the coherent variant additionally requires deep block-encoding circuits and non-trivial state preparation, and the SVD pre-processing itself carries a non-negligible classical overhead. No single encoding dominated across problem sizes, which motivates continued exploration of compressed and structure-aware encodings.

\textbf{Ansatz and optimizer.} Across 11 ansatz families on Hele--Shaw flow and sample-tridiagonal Toplitz matrix, we found that expressibility and entanglement metrics correlate only weakly with VQLS convergence, so off-the-shelf circuit metrics are not a reliable proxy for solver quality. Among classical optimizers, gradient-based methods incurred a higher per-iteration cost than gradient-free alternatives but converged more reliably in the HPC setting we tested. Together, these observations argue for problem-aware ansatz design rather than ansatz selection by generic expressibility scores.

\textbf{HPC deployment.} We deployed the end-to-end workflow on the OLCF Frontier supercomputer and successfully simulated a 15-qubit tridiagonal Toeplitz system on a single node, demonstrating a increasing performance with proposed optimization algorithms.

Our ongoing work extends the analysis to higher-complexity CFD problems, real quantum-hardware execution alongside the HPC simulator, problem-aware and physics-informed ansatz construction, and additional matrix-encoding strategies aimed at further reducing LCU term count and circuit depth. We see these as the immediate steps toward a deployable hybrid quantum--HPC linear-solver stack for fluid dynamics.

%% file: Sections/Acknowledgments.tex
\section{Acknowledgments}
The authors would like to thank Pooja Rao for her valuable contributions and insightful discussions that helped shape this work. 

This research used resources of the Oak Ridge Leadership Computing Facility at the Oak Ridge National Laboratory, which is supported by the Office of Science of the U.S. Department of Energy under Contract No. DE-AC05-00OR22725.

\section{Acknowledgement of AI-Generated Content}
Portions of this manuscript were prepared with the assistance of large language model (LLM) tools, which were used to support drafting, editing, grammar refinement, and stylistic improvements of the text. All AI-generated content was reviewed, verified, and edited by the authors, who take full responsibility for the content in this work. No AI tools were used to generate or fabricate experimental data, numerical results, figures, or citations; all references were independently verified by the authors.